\documentclass[sigconf]{acmart}
\makeatletter
\def\@ACM@checkaffil{
    \if@ACM@instpresent\else
    \ClassWarningNoLine{\@classname}{No institution present for an affiliation}%
    \fi
    \if@ACM@citypresent\else
    \ClassWarningNoLine{\@classname}{No city present for an affiliation}%
    \fi
    \if@ACM@countrypresent\else
        \ClassWarningNoLine{\@classname}{No country present for an affiliation}%
    \fi
}
\makeatother

\AtBeginDocument{%
  \providecommand\BibTeX{{%
    \normalfont B\kern-0.5em{\scshape i\kern-0.25em b}\kern-0.8em\TeX}}}

\usepackage{amsmath,amsfonts}
\usepackage{algorithmic}
\usepackage{graphicx}
\usepackage{textcomp}
\usepackage{xcolor}
\usepackage{balance}
\usepackage{natbib}
\usepackage{subfigure}
\usepackage{geometry}

\copyrightyear{2023} 
\acmYear{2023} 
\setcopyright{acmlicensed}\acmConference[WWW '23 Companion]{Companion Proceedings of the ACM Web Conference 2023}{April 30-May 4, 2023}{Austin, TX, USA}
\acmBooktitle{Companion Proceedings of the ACM Web Conference 2023 (WWW '23 Companion), April 30-May 4, 2023, Austin, TX, USA}
\acmPrice{15.00}
\acmDOI{10.1145/3543873.3584638}
\acmISBN{978-1-4503-9419-2/23/04}

\begin{document}

\title{Learning Multi-Stage Multi-Grained Semantic Embeddings for E-Commerce Search}

\author{Binbin Wang\textsuperscript{\rm *}, Mingming Li, Zhixiong Zeng, Jingwei Zhuo, Songlin Wang, Sulong Xu, Bo Long, Weipeng Yan}

\email{{wangbinbin77,limingming65, zengzhixiong7, zhuojingwei1,wangsonglin3,xusulong, bo.long, Paul.yan}@jd.com}
\affiliation{
  \institution{JD.com, Beijing, China}
}


\begin{abstract} 
Retrieving\let\thefootnote\relax\footnotetext{* Corresponding author.} relevant items that match users' queries from billion-scale corpus forms the core of industrial e-commerce search systems, in which embedding-based retrieval (EBR) methods are prevailing.
These methods adopt a two-tower framework to learn embedding vectors for query and item separately and thus leverage efficient approximate nearest neighbor (ANN) search to retrieve relevant items.
However, existing EBR methods usually ignore inconsistent user behaviors in industrial multi-stage search systems, resulting in insufficient retrieval efficiency with a low commercial return. 
To tackle this challenge, we propose to improve EBR methods by learning Multi-level Multi-Grained Semantic Embeddings (\textbf{MMSE}). We propose the multi-stage information mining to exploit the ordered, clicked, unclicked and random sampled items in practical user behavior data, and then capture query-item similarity via a post-fusion strategy. We then propose multi-grained learning objectives that integrate the retrieval loss with global comparison ability and the ranking loss with local comparison ability to generate semantic embeddings.
Both experiments on a real-world billion-scale dataset and online A/B tests verify the effectiveness of MMSE in achieving significant performance improvements on metrics such as offline recall and online conversion rate (CVR).

\end{abstract}



\maketitle

\section{Introduction}
E-commerce search system has been an important technique to satisfy the emergent needs of online shopping.
It is a essential part of e-commerce shopping platforms (\emph{e.g.},  Ebay, Amazon, Taobao, JD and so forth), which serves for hundreds of millions of daily active users and contributes to the largest percentage of transactions among all channels \cite{zhang2020towards,sondhi2018taxonomy,liu2017cascade}. 
Existing search system mainly adopts the multi-stage architecture of ``retrieval-preranking-ranking'' to discover and screen items that satisfy users.
The retrieval stage plays a important role in determining the quality of the candidate set in the downstream ranking stages \cite{li2021embedding}, thus receiving salient attention from both academia and industry.

The challenge of the retrieval stage is to find relevant items from billion-level corpus in low latency and computational cost.  The most represented work is the embedding-based retrieval (EBR)  \cite{huang2013learning,grbovic2018real,huang2020embedding,wu2020zero,liu2021pre}, which adopts the paradigm of building two-tower EBR system to provide relevant or personalized items \cite{li2021embedding,zhang2020towards,nigam2019semantic,lu2022ernie,khattab2020colbert,santhanam2021colbertv2,haldar2020improving,qu2020rocketqa, karpukhin2020dense,luan2021sparse,li2022learning,lindgren2021efficient}. 



However, existing multi-stage search systems usually model inconsistent user behaviors at different stages, which means that the candidate items obtained in the retrieval stage cannot effectively serve following the ranking stage.
Specifically, the retrieval phase focuses on the interactions with click or exposed items and sampled negative items, while ignores the interactions with ordered items and unclicked items (\emph{i.e.}, exposed but unclicked) that are consistent with the following ranking stage and more relevant to business indicators (\emph{i.e.}, CTR, CVR). 
In fact, due to the problems of multi-level similarity and inconsistent sample distributions, it is nontrivial to integrate multi-stage information in the retrieval stage simultaneously:
For example, the ordered item is obviously more relevant to the query than the clicked item in the user behaviors, yet existing retrieval algorithms are ineffective to handle the multi-level similarity problem in the embedding space. Therefore, the development and rational integration of multi-stage information is a challenging problem, leading to great potential in improving the consistency of multi-stage search systems as well as the commercial return.


\begin{figure*}[tbp]

\begin{minipage}[t]{0.5\textwidth}
\centering
\includegraphics[width=9cm]{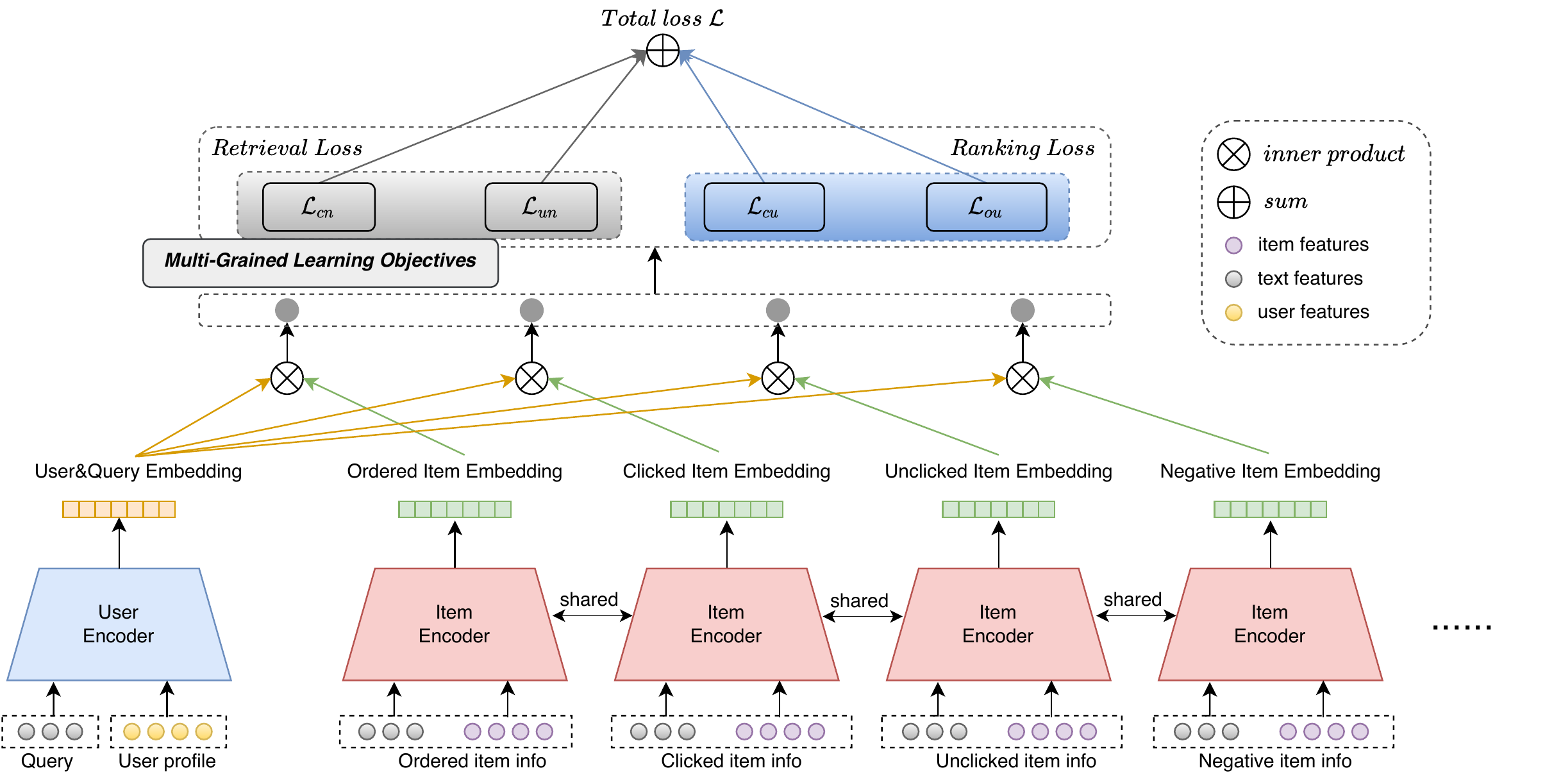}
\caption{Architecture of MMSE}
 \label{fig:model}
\end{minipage}
\begin{minipage}[t]{0.4\textwidth}
\centering
\includegraphics[width=5cm]{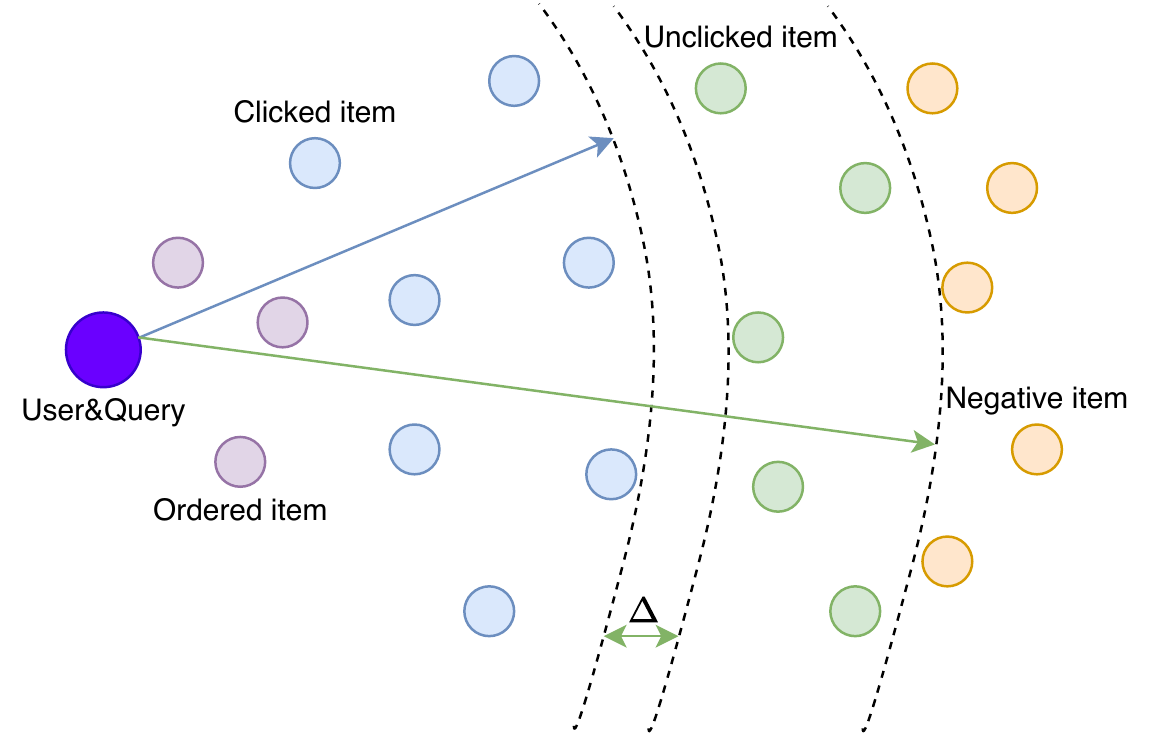} 
\caption{Illustration of Multi-Grained Learning}
\label{multiloss}
\end{minipage}
\end{figure*}

   

To tackle these issues, in this paper, we propose to learn Multi-stage Multi-grained Semantic Embeddings (\textbf{MMSE}) for e-commerce search. 
Specifically, we propose a multi-stage information mining framework to exploit both ordered, clicked and exposed but unclicked items, which employs a post-fusion strategy to project multi-stage items into a shared item space.
To preserve the multi-level similarity, we design multi-grained learning objectives to encourage a rank of similarities between multi-stage items. It generally includes two types of losses, the retrieval loss with global comparison ability for retrieval efficiency, and the ranking loss with local comparison ability for multi-stage integration.
We conduct experiments on a real e-commerce dataset of one billion scale collected from JD.com, and show that experimental results have achieved significant improvements in offline performance metrics. In addition, MMSE also achieves significant performance improvement in online A/B testing. It is worth mentioned that the introduction of multi-level semantic similarity also improves the final conversion rate (CVR).

Our main contributions are summarized as follows:
\begin{itemize}
    \item We propose multi-stage information mining, which to our best knowledge is the first work to develop multi-stage items in the retrieval model to facilitate the consistency of the e-commerce search system.
    \item We propose multi-grained learning objectives to integrate multi-stage items effectively via deploying retrieval loss with global comparison ability and ranking loss with local comparison ability.
    \item Experiments conducted on a billion-level industrial dataset and online A/B test demonstrate the effectiveness of MMSE.
\end{itemize}

\section{Proposed Method}


Figure ~\ref{fig:model} illustrates the overall architecture of the multi-stage multi-grained semantic embedding method (MMSE). We will introduce the design of model architecture, the multi-stage information mining,  and multi-grained learning objectives in details.

\subsection{Two-Tower Model Architecture}
The typical solution for e-commerce retrieval adopts the two-tower model architecture, which consists of a query tower $f_q(\cdot)$ and an item tower $f_p(\cdot)$ to process query and item features.  Typically, the query consists of both text feature and user profile feature, and the item contains both text feature and item feature.
Given the query feature $x_q$ and the item feature $x_p$, the model outputs the query embedding and item embedding by:
\begin{equation}
\textbf{q} = f_q(x_q, \theta_q) \in \mathbb{R}^d, \quad \textbf{p} = f_p(x_p;\theta_p) \in \mathbb{R}^d\label{eq1}
\end{equation}
where $d$ denotes the dimension of embedding, $\theta_q$ and $\theta_p$ are learnable parameters of the two-tower model. Here both $f_q(\cdot)$ and $f_p(\cdot)$ are 4-layer Transformer pre-trained on e-commerce data.
Then we can deploy a scoring function $S(q, p)$ to compute the similarity between query and item, denoted as:
\begin{equation}
S(q, p) = \left<\textbf{q}, \textbf{p} \right>\label{eq2}
\end{equation}
where $\left< ,\right>$ denotes the inner product.
The core design of the two-tower model is to learn the query and item embeddings separately, thereby facilitating the offline computation of item embeddings to improve retrieval efficiency.

\subsection{Multi-stage Information Mining}

To ensure multi-stage information consistency and improve business returns, we propose multi-stage information mining to train retrieval models using ordered, clicked and unclicked items simultaneously.
We adopt a post-fusion strategy to project these items into a shared embedding space via the item tower $f_p(\cdot)$, and fuse them by designing multi-grained learning objectives.
Formally, given a user and his query ${q}$, 
the embedding set of user ordered items ${P}_o$,
 the embedding set of user clicked items ${P}_c$, 
 and the set of user unclicked items ${P}_u$. Note that ${P}_o$ is a small subset of ${P}_c$, i.e., $P_o \subset P_c$, since the ordered behavior is usually sparse.
We employ a hybrid approach to select two sources of negative samples and generate the negative's set ${P}_n$, including the random negatives sampled from all candidate items and the batch negatives collected by permuting the positive query-item pairs in a training batch. What's more, 
we utilize the shared random negative samples for all training examples in a training batch, which is empirically efficient and reduces the computational cost.
Finally, for each user query $q$, the corresponding multi-stage information can be formulated as ${P}=[{P}_o,{P}_c, {P}_u,{P}_n]$.

\subsection{Multi-Grained Learning Objectives}

Figure \ref{multiloss}  illustrates the idea of designing multi-grained learning objectives. In the following content, we will elaborate in detail.

\subsubsection{Metric of clicked-negative pairs}

As the goal of retrieval is to find the top@$K$ items relevant to the query from all candidates, which requires sufficient global comparison ability. Therefore, the most common retrieval loss is the softmax cross-entropy loss, which takes the clicked items $P_c$ and the negative items $P_n$ as input to calculate the loss by:
\begin{equation}
\mathbf{\mathcal{L}}_{cn} = -\sum_{p_c\in P_c}\log\frac{\exp(S(q,p_c)/\tau_1)}{  \sum_{p_i\in P_n \cup \{p_c \}}\exp(S(q,p_i)/\tau_1)}\label{eq3}
\end{equation}
where $\tau_1$ is a temperature parameter to smooth the overall fitted distribution of the training data. 

\subsubsection{Metric of unclicked-negative pairs}
Previous work ignores the order information $P_o$ and unclicked information ${P}_u$ that is critical for downstream ranking tasks and commercial returns. However, it is challenging to utilize ${P}_o$ and ${P}_u$ in a reasonable way, due to the newly introduced multi-level similarity problem. Specifically, we cannot simply treat the ${P}_o$ and ${P}_u$ as positive or negative samples to apply the common training objective in \eqref{eq3}.
Therefore, we consider multi-grained modeling for multi-stage information $P$. Considering the unclicked item ${P}_u$, it is more similar to the query than the negative item ${P}_n$, but not as similar as the clicked item ${P}_c$. We accordingly present two learning objectives to preserve these two similar relations. For the relation between ${P}_u$ and ${P}_n$, we propose the unclicked-negative retrieval loss to ensure the global comparison ability of the negative sample space:
\begin{equation}
\mathbf{\mathcal{L}}_{un} = -\sum_{p_u\in P_u}\log\frac{\exp(S(q,p_u)/\tau_2)}{ \sum_{p_i\in P_n \cup \{p_u\}}\exp\left(S(q,p_i)/\tau_2\right)}\label{eq4}
\end{equation}
where $\tau_2$ is similar to $\tau_1$, denotes the temperature parameter. 

\subsubsection{Metric of clicked-unclicked pairs}

 For the relation between ${P}_c$ and ${P}_u$, we propose a ranking loss to preserve their local relations.
we select the hinge loss to encourage their similarity difference greater than a margin, which can be formulated as:
\begin{equation}
\mathbf{\mathcal{L}}_{cu} = \sum_{p_u\in P_u}\sum_{p_c\in P_c}\left[S(q,p_u) - S(q,p_c) + \Delta\right]_+ \label{eq5}
\end{equation}
where $[x]_+$ denotes the $\max(x,0)$,  $\Delta$ denots the margin for similarity separation.

\subsubsection{Metric of ordered-unclicked pairs}
To preserve the multi-level similarity of ordered items ${P}_o$, we need to carefully consider its sparse characteristic and the subset relationship between ${P}_o$ and clicked items ${P}_c$. 
Noting that, it is unable to construct a ranking loss for ${P}_o$ and ${P}_c$ due to their subset relationship, and also inappropriate for ${P}_o$ and ${P}_n$ due to the extreme sparsity of positive samples. 
We empirically find that modeling a ranking loss for ${P}_o$ and ${P}_u$ is beneficial, and propose a ordered-unclicked bayesian personalized ranking loss to maximize their similarity difference:
\begin{equation} 
\mathbf{\mathcal{L}}_{ou} = -\sum_{p_o\in P_o}\sum_{p_u\in P_u}\log\sigma(S(q,p_o) - S(q,p_u))\label{eq6}
\end{equation}
where $\sigma(\cdot)$ denotes the sigmoid activation function. $\mathbf{\mathcal{L}}_{ou}$ is more suitable for modeling sparse ordered data, as it avoids complex tuning of parameters like hinge loss.

Finally, combining the modeling of multi-level similarity provided by \eqref{eq3} - \eqref{eq6}, the proposed multi-grained learning objective is:
\begin{equation}
\mathbf{\mathcal{L}}= {\mathcal{L}_{cn}}+{\mathcal{L}_{un}}+{\mathcal{L}_{cu}}+{\mathcal{L}_{ou}}\label{eq7}
\end{equation}

\begin{table*}[htbp]
    \centering
    \caption{Offline experimental results on LTR v.s. MMSE in terms of batch-top@1, recall@K (K is set to 1, 50, 100, 500, 1000)}.
    \label{tab:main_results1}

    \begin{tabular}{c|c|c|c|c|c|c}
    \toprule
        Methods & batch-top@1 & recall@1 & recall@50 & recall@100 & recall@500 & recall@1000 \\
     \midrule
        DSSM$^{\mathrm{a}}$ \cite{huang2013learning} & 0.9046 & 0.0069 &  0.1789 & 0.2759 & 0.5706 & 0.6806 \\
    \textbf{DSSM+MMSE} & \textbf{0.9278} & \textbf{0.0228} & \textbf{0.4063} & \textbf{0.5285} & \textbf{0.7517} & \textbf{0.8067} \\
    \midrule
    
    \midrule
        DPSR \cite{dpsr} & 0.9055 & 0.0076 & 0.1946 & 0.2939 & 0.5771 & 0.6817 \\
        \textbf{DPSR+MMSE} & \textbf{0.9287} & \textbf{0.0237} & \textbf{0.3993} & \textbf{0.5218} & \textbf{0.7447} & \textbf{0.8014} \\
    \midrule
    \midrule
        LTR \cite{liu2009learning}& 0.8569 & 0.0061 & 0.1558 & 0.2333 & 0.4782 & 0.5835 \\
    \midrule
        RSR \cite{qiu2022pre} & 0.8898 & 0.0094 & 0.1980 & 0.2888 & 0.5486 & 0.6496 \\
        RSR+${M_1}$  ($\mathbf{\mathcal{L}}_{cn}+\mathbf{\mathcal{L}}_{un}$) & 0.9245 & 0.0076 & 0.1932 & 0.2909 & 0.5622 & 0.6583 \\
        RSR+${M_2}$ ($\mathbf{\mathcal{L}}_{cn}+\mathbf{\mathcal{L}}_{un}+ \mathbf{\mathcal{L}}_{cu}$) & 0.9270 & \textbf{0.0107} & 0.2108 & 0.3072 & 0.6025 & 0.6993 \\
        \textbf{RSR+MMSE} & \textbf{0.9303} & 0.0099 & \textbf{0.2201} & \textbf{0.3267} & \textbf{0.6145} & \textbf{0.7104} \\
  
    \bottomrule        
    \multicolumn{7}{l}{$^{\mathrm{a}}$The vocabulary size and batch size of DSSM/DPSR is set to 400k, 1024, while RSR is 20k, 350.}
   
    \end{tabular}
    
\end{table*}


\section{Experiments}
In this section, we describe the experimental datasets, metrics, baselines,
and specific experimental setup. 
After that, we make various ablation studies to investigate the  affect of unclicked/ordered data and target features for dense retrieval.

\subsection{Experimental Protocols}

\subsubsection{Datasets and Metrics}
We conduct extensive experiments on real-world E-commerce scenario, which contains user purchasing and clicking behaviors.  We collect search logs of the user for over two months from JD E-commerce website,  where the size of the dataset exceeds one billion, and the number of users and items is a hundred millions
scale.

To evaluate the offline performance and quality,
we adopt two widely used metrics for offline \cite{liu2021que2search,li2021embedding}: batch-top@1, and recall@K ($K \in\{1, 50, 100, 500, 1000\}$. 
To evaluate the online performance, we choose several retrieval metrics, such as UV-value (revenue per Unique Visitor), RPM (Revenue Per Mile), and CVR (Click Value Rate), to measure the results of the A/B test. 
we also measure the performance by the number of items after passing the relevance module,  participating in the preranking phase, denoted as ${Num_{prank}}$.


\subsubsection{Baselines}
To verify the effectiveness of MMSE, we compare it with the classical method of  Learning to Rank (LTR) \cite{liu2009learning}, DSSM \cite{huang2013learning}, DPSR \cite{dpsr}, RSR (base) \cite{qiu2022pre}, ours and various variants RSR+$M_*$, such as ${M_1}=\mathcal{L}_{cn} + \mathcal{L}_{un}$, ${M_2}=\mathcal{L}_{cn} + \mathcal{L}_{un} +\mathcal{L}_{cu}$.


\subsubsection{Implementation Details}
To ensure a fair comparison among
different methods, we keep the feature columns, vocabulary
size, the dimension of query/item, and parameters of product quantization (PQ) \cite{ge2013optimized}
unchanged. Specifically, we set the dimension as 128, batch
size as 350, n-list of IVF-PQ as 32768, and nprobe is set to 32. $\tau_1$ and $\tau_2$ of softmax are set to 1/30. The margin $\Delta$ is set to 0.02.  The AdamW optimizer \cite{loshchilov2017decoupled} is employed with an initial learning rate of 5e-5. The maximum length of query and item title sequences are 30 and 100, respectively. Moreover, the indexing construction is based on the Faiss ANNS library\footnote{https://github.com/facebookresearch/faiss}. 

\begin{table}[tbp]
    \centering
    \caption{Online performance of A/B tests. The improvements are
averaged over 10 days in May 2022. p-value is obtained by t-test over the RSR.}
    \label{tab:ab_test}
    {
    \begin{tabular}{c|c|c|c|c}
    \toprule
         & ${Num_{prank}}$  & RPM & CVR & UV-value \\
    \midrule
        Gain & +2.0\% & +0.35\% & +0.30\% &  +0.47\% \\
        p-value$^{\mathrm{b}}$ & - & 0.0013 & 0.0004 & 0.0866 \\
    \bottomrule
    \multicolumn{4}{l}{$^{\mathrm{b}}$  Small p-value means statistically significant.}
    \end{tabular}
    }
\end{table}

\subsection{Main Results}

\subsubsection{Offline results}
Table \ref{tab:main_results1} shows the overall offline performances of our proposed
method.  We can summarize some insights as follows: 1) Comparing with baselines, we can observe that MMSE achieves a significant improvement with most  evaluation metrics, illustrating  the superiority and effectiveness.  2) Compared  with LTR,  MMSE could achieve better performance, which explains that the metric of clicked and ordered is useful to explore the fine-grained relationships. 3) Comparing with base and variants, we can see that multi-stage  information plays an important role in dense retrieval, the more information and relationships measured, the better performance obtained, implying the consistency with business goals. 4) Comparing with variants, we find that the method of train with unclicked data has a slight improvement than base, while train with unclicked + clicked  brings substantial improvements, revealing that unclicked data is useful, while it is hard to learn, and our designed metric loss of unclicked and clicked benefits the representation learning process.

\subsubsection{Online results}

To investigate the effectiveness of the proposed method in the real-world commercial scenario, we conduct several A/B tests, and the online results are reported in Table \ref{tab:ab_test}.  Comparing with the base model in the real online environment, we can note that our performance of MMSE increases by 2.0\% in terms of $Num_{prank}$ and 0.30\% in CVR, respectively,  which demonstrates that the designed techniques are practical gains for the online system. 
\subsection{Performance in ordered datasets}

\begin{figure}[tbp]
    \centering
    \includegraphics[scale=0.45]{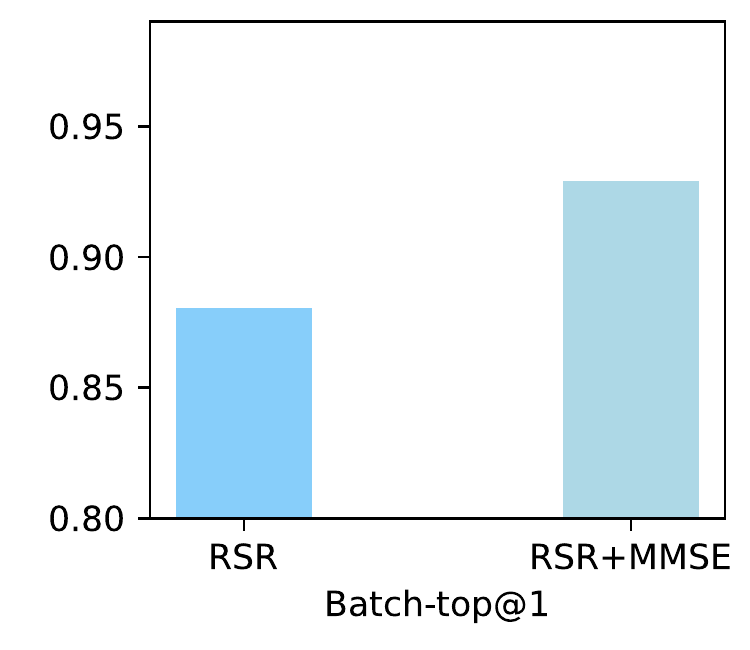} 
    \includegraphics[scale=0.45]{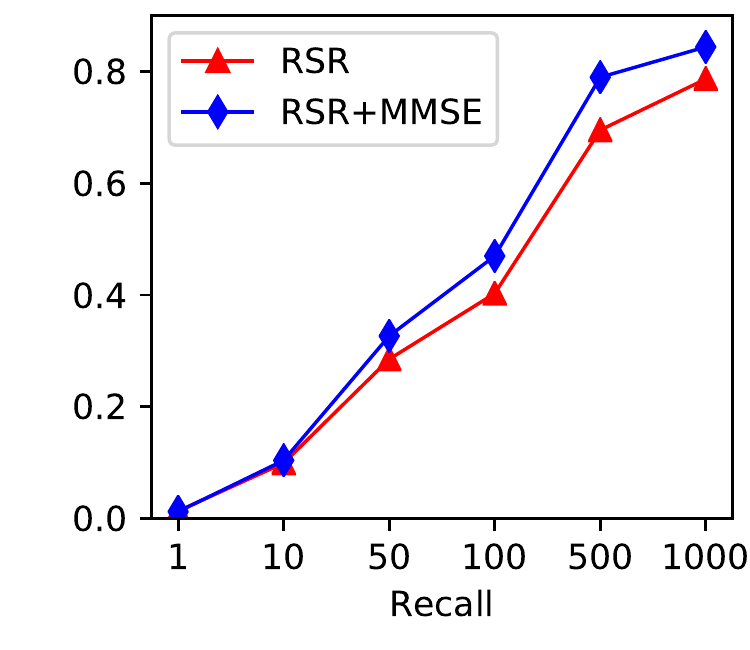}
    \caption{Batch-top@1 (left) and recall (right) in ordered datasets}
    \label{fig:ordered_recall}
\end{figure}
\begin{table}[tbp]
    \centering
    \caption{An ablation study on features. w/o denotes the method without additional features, and w denotes additional order related features is used.}
    \label{tab:features}
    {
    \begin{tabular}{c|c|c|c}
    \toprule
        Methods & batch-top@1 & recall@1 &   recall@1000 \\
    \midrule
        RSR (w/o) & 0.8898 & 0.0094     & 0.6496 \\
        RSR (w) & 0.9054 & \textbf{0.0132}   & 0.6932 \\
        RSR+MMSE (w/o) & \textbf{0.9303} & {0.0099}   & \textbf{0.7104} \\
    \bottomrule
    \end{tabular}
    }
\end{table}
To verify the role of ordered data in retrieval stage, we conduct extensive experiments to evaluate the performance of RSR and RSR+MMSE in ordered datasets. 
The result is reported in Figure \ref{fig:ordered_recall}. Obviously, our proposed method outperform RSR significantly on various evaluation metrics. This results show that the ordered data is useful for metric learning process, measuring the relationship of ordered-unclicked pairs simultaneously. The ordered data is sparse but with stronger signal than clicked data, which will push the unclicked data away from the query and further distinguish the several clicked data well.

\subsection{Ablation study on features}
According to the above analysis, we can find that ordered data is useful for ordered-unclicked pair learning from the perspective of optimization. Here, we would like to investigate the effect of order features, such as sales volume, average rating, comments number, and so on. Table \ref{tab:features} reports the main results. 

Compared with the RSR without order features (RSR (w/o)), the RSR with order features (RSR (w))
obtains a higher recall, indicating that additional features of orders provide high-quality guidance regarding dense retrieval. RSR+MMSE outperforms the RSR(w) by a large margin in almost all different metrics, implying that optimization of metric learning brings greater benefits than feature integrated remarkably. In summary, the ordered features will bring a great improvement, and multi-grained learning objectives designed will make it be fully exploited.

\section{Conclusion}

In this paper, we identify the problem of inconsistent information that hinder the improvement of the e-commerce search system. 
To tackle this issue, we propose to learn Multi-level Multi-Grained Semantic Embeddings (MMSE), consisting of multi-stage information mining to develop multi-stage consistent items and multi-grained learning objectives to integrate these items effectively. Offline and online experiments conducted on a billion-level industrial dataset demonstrate the effectiveness of our proposed method.

\bibliographystyle{ACM-Reference-Format}
\bibliography{./cited.bib}

\end{document}